\documentclass{article}
\usepackage{lipsum}
\usepackage{caption}
\usepackage{subcaption}
\usepackage{authblk}
\usepackage[english]{babel}
\usepackage{amsfonts}
\usepackage{amsmath}
\usepackage[top=2cm, bottom=2cm, left=2cm, right=2cm]{geometry}
\usepackage{fancyhdr}
\usepackage{multicol}
\usepackage{bbold}
\usepackage{xcolor}
\usepackage{epsfig}
\usepackage{braket}

\usepackage{graphics}
\usepackage{epsfig}
\usepackage{lipsum}
\usepackage{fancyhdr}
\usepackage{booktabs}
\usepackage{cleveref}
\usepackage{wrapfig}
\usepackage[export]{adjustbox}
\usepackage{rotating}

\renewenvironment{abstract}{
	
	\hfill\begin{minipage}{0.95\textwidth}
		\rule{\textwidth}{1pt}}
	{\par\noindent\rule{\textwidth}{1pt}\end{minipage}
}



\begin{document}
	\title{\textbf{Optimal Superdense Coding Capacity in the Non-Markovian Regime}}
	\author[1]{\textbf{Y. Aiache}}
	\author[3]{\textbf{S. Al-Kuwari}}
	\author[1]{\textbf{K. El Anouz}}
	\author[1,2]{\textbf{A. El Allati}}
	\affil[1]{\small Laboratory of R\&D in Engineering Sciences, Faculty of Sciences and Techniques Al-Hoceima,\newline Abdelmalek Essaadi University, Tetouan,
		Morocco}
	\affil[1]{\small Max Planck Institute for the Physics of Complex Systems, Nöthnitzer Str. 38, D-01187 Dresden, Germany}
	\affil[3]{\small Qatar Center for Quantum Computing (QC2), College of Science and Engineering, \newline Hamad Bin Khalifa University, Doha, Qatar}
	\maketitle
	\begin{center}
		\textbf{Abstract}
	\end{center}
	\begin{abstract}
		Superdense coding is a significant technique widely used in quantum information processing. Indeed, it consists of sending two bits of classical information using a single qubit, leading to faster and more efficient quantum communication. In this paper, we propose a model to evaluate the effect of backflow information in a superdense coding protocol through a non-Markovian dynamics. The model considers a qubit interacting with a structured Markovian environment. In order to generate a non-Markovian dynamic, an auxiliary qubit contacts a Markovian reservoir in such a way that the non-Markovian regime can be induced. By varying the coupling strength between the central qubit and the auxiliary qubit, the two dynamical regimes can be switched interchangeably. An enhancement in non-Markovian effects corresponds to an increase in this coupling strength. Furthermore, we conduct an examination of various parameters, namely temperature weight, and decoherence parameters in order to explore the behaviors of superdense coding, quantum Fisher information, and local quantum uncertainty using an exact calculation. The obtained results show a significant relationship between non-classical correlations and quantum Fisher information since they behave similarly, allowing them to detect what is beyond entanglement. In addition, the presence of non-classical correlations enables us to detect the optimal superdense coding capacity in a non-Markovian regime. 
	\end{abstract}
	
	\section{Introduction}
	\label{intro}
	
	In the theory of open quantum systems, the interaction between quantum systems and their surrounding environment leads to the phenomenon of decoherence \cite{Ref1,Ref2}. Consequently, the implementation of quantum protocols presents a significant challenge because of the exchange of information during the interaction between the system and its environment. In this context, the dynamics of the open system cannot be described by unitary evolution. To solve this problem, several methods were explored, including quantum dynamical maps \cite{Ref3}, and the quantum master equation \cite{Ref4}. However, irrespective of how the open system and its environment interact, certain general features of the reduced system dynamics can be evaluated. In a Markovian process, an open system continuously loses information about its surroundings, whereas non-Markovian processes exhibit an information flow from the environment back to the open system \cite{Found}. This shows that the existence of memory effects serves as an essential characteristic of non-Markovian quantum behavior.\\
	
	In the field of quantum information science, entanglement is defined as a special kind of quantum correlation that plays a crucial role in the implementation and manipulation of protocols such as quantum cryptography and quantum teleportation \cite{Ref5,Ref6,Ref7,Ref.6,Ref9,khadija 3}. However, recent research has shown that quantum correlations extend beyond entanglement \cite{Ref10}. Furthermore, non-classical correlations have been demonstrated for certain separable quantum states, which can arise without entanglement \cite{Ref11}. Moreover, several tools have been proposed to quantify the amount of non-classical correlations. In particular, Girolami \emph{ et al.} \cite{Girolami} introduced the concept of Local Quantum Uncertainty (LQU) as a measure of non-classical correlations. In fact, this measure quantifies the uncertainty that arises from the measurement of a single local observable in a quantum state \cite{khadija 1}. Furthermore,  many techniques in quantum metrology explore the application of quantum mechanics principles to achieve highly precise measurements, surpassing the limitations of classical methods \cite{Giovannetti}. \\

	One of the most important concepts in quantum metrology is Quantum Fisher Information (QFI) \cite{allati12, appl}, which has been widely used in various fields, including quantum phase transition \cite{Ref12}, quantum speed limit time \cite{Ref13}, uncertainty relation \cite{Ref14}.  Quantum Fisher Information (QFI) is a measure of the information carried by a measurement result on an unknown parameter \cite{Fisher}. It also uses statistical properties to investigate the measurement of the precision limit of the estimation parameters. Moreover, it serves to establish the link between quantum correlations and quantum metrology \cite{KE,Ref15,Ref16}, and attempts to characterize the highest achievable precision in parameter estimation protocol \cite{khadija 2}. In fact, the QFI satisfies the monotonicity criterion and can detect non-Markovian behavior during the interaction between an open system and its environment \cite{Ref17}.\\
	
	In this paper, our aim is to investigate exactly the dynamics of superdense coding (SDC), QFI, and LQU under non-Markovian regimes. We attempt to answer the question of whether it is possible to detect optimal superdense coding through the dynamics of non-classical correlations and non-Markovianity. We will quantify non-Markovianity based on the notion of trace distance \cite{traced}, where the trace distance is used as a quantifier tool to distinguish between Markovian and non-Markovian behaviors. To achieve this, it is assumed that a qubit is coupled to a structured non-Markovian environment, and the environment is divided into two main parts: another qubit and a Markovian reservoir. \\
	
	The rest of this paper is structured as follows: in Sec.(\ref{sec:3}), we present a brief review of some preliminaries, such as SDC, QFI, LQU and non-Markovianity. In Sec.(\ref{sec:2}) introduces the Hamiltonian that describes the proposed model and solves the master equation of the suggested open system, namely the composite qubit-qubit system. Sec.(\ref{sec:4}) contains a numerical discussion of our results. Finally, in Sec.(\ref{sec:5}), a summary of our results and concluding remarks are provided.

	\section{Preliminaries}
	\label{sec:3}
	
	In the this section, we aim to define some preliminary concepts such as superdense coding, quantum Fisher information, local quantum uncertainty and trace distance. 
	\subsection{Superdense coding protocol}
	
	Superdense coding is a fascinating non-classical implication. Roughly speaking, by mixing two qubits maximally entangled, the sender can transmit two bits of classical information to the receiver by transmitting a single qubit \cite{DC1}. However, the concept of superdense coding is extremely important in information theory. For example, superdense coding uses correlated states to transmit and manipulate quantum teleportation \cite{DC2}.\\
	
\begin{figure}[h]
		\centering
		\resizebox{0.6\hsize}{!}{\includegraphics*{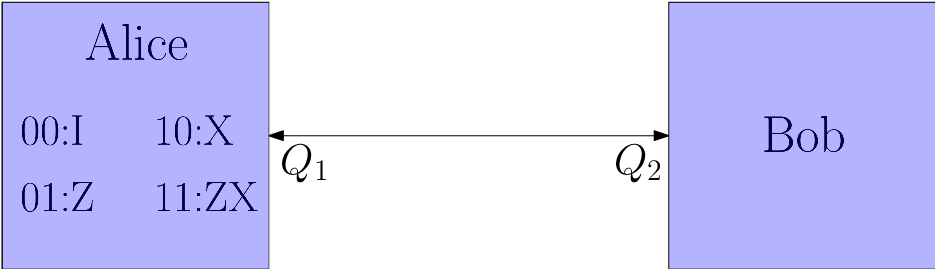}}
		\caption{The basic setup for superdense coding involves Alice and Bob, each holding one part of a correlated pair of qubits. When Alice sends her single qubit to Bob, she can effectively transmit two bits of classical information to him using the process known as superdense coding. The procedure is: when sending a string of bits to Bob, if she wants to send '00', she leaves her qubit unchanged. For '01', she performs the phase flip $Z$ to her qubit. To send '10', she applies the quantum gate $X$. Finally, for '11', she applies the $ZX$ gate to her qubit.}
		\label{tt}
	\end{figure}

	In our case, we perform superdense coding in a composite system consisting of two qubits, namely $ Q_1 $ and $ Q_2 $. To achieve this, we can use a set of mutually orthogonal unitary transformations (MOUT). In fact, the MOUT for any two-qubit system is given in the following form \cite{Ref19}:
	\begin{equation}
		\mathcal{U}_{i=(pq)}\ket{j}=e^{\sqrt{-1}\; \pi\;p \;j}\ket{j+q(\hbox{Mod}(2))},
	\end{equation}
	where $ \ket{j} $ is the computational basis of the qubit ($ \ket{j}=\{\ket{0},\ket{1}\} $) and $\{pq\}\equiv\{00,01,10,11\}$. The average state of the signal set after these unitary transformations ($\mathcal{U}_{i}$) becomes
	\begin{equation}\label{Average}
		\overline{\rho_{Q}^*}(t)=\dfrac{1}{4}\sum_{i=0}^{3}
		\big( \mathcal{U}_{i}\otimes \mathcal{I}_{2}  \big) \rho_{Q}(t)
		\big( \mathcal{U}_{i}^\dagger\otimes \mathcal{I}_{2}  \big),
	\end{equation}
	where $ \mathcal{I}_{2}$ denotes the identity operator in 2-dimension. While $ i=0 $ holds for $ (pq)=(00) $, $i=1$ stands for $(01)$, while $i=2$ and $i=3$ for $(10)$ and $(11)$ respectively. However, one can set the maximum amount of classical information that could be communicated to a receiver for a defined initial state, called the superdense coding capacity (see Fig.(\ref{tt})). Hence, in the case where the sender, Alice, performs the set of MOUT, the maximum superdense coding capacity can be expressed as \cite{Ref19}
	\begin{equation}
		SDC(t)=\mathcal{S}(\overline{\rho_{Q}^*}(t))-\mathcal{S}(\rho_{Q}(t)),
	\end{equation}
	where $ \mathcal{S}(\overline{\rho_{Q}^*}(t)) $ is the von-Neumann entropy of the average state of the set of signal states $ \overline{\rho_{Q}^*}(t)$, while $ \mathcal{S}(\rho_{Q}(t)) $ is the von-Neumann entropy of the density matrix $ \rho_{Q}(t) $. Moreover, the superdense coding is valid if $ SDC>1 $. However, the optimal superdense coding corresponds to the maximum value, i.e., $ SDC=2 $. The latter condition means that the sender can transmit two bits of classical information by sending one qubit.\\
	
	Superdense coding is a fundamental concept in quantum information processing that promises efficient communication and has been extensively studied both theoretically and experimentally \cite{Mattle}. Its applications extend to various quantum technologies and inspire research in quantum communication and computation. In our current study, we aim to show if there is a relationship between superdense coding and quantum metrology using the  quantum Fisher information that will be presented in the next part.\\
	
	\subsection{Quantum Fisher information}
	\label{sub:2}
	Quantum Fisher information is a measure of the amount of information that can be extracted from a quantum state. It also quantifies the sensitivity and precision of the quantum measurement of the state \cite{Ref20}.  QFI can be seen as a measure of the flow of information between an open system and its surrounding environment \cite{Ref17}. For a given density matrix $ \rho_{Q}(t) $, QFI is defined as \cite{Ref22}
	\begin{equation}
		\mathcal{F}(\rho_{Q}(t),H^{Q})=\dfrac{1}{4}\hbox{Tr}\{\rho_{Q}(t)L^2\},
	\end{equation}
	where $ L $ is the symmetric logarithmic derivative operator, which satisfies the following equation 
	\begin{equation}
		\dfrac{d\rho_{Q}(t)}{dt}=\dfrac{1}{2}\big( L\rho_{Q}(t)+\rho_{Q}(t)L \big).
	\end{equation}
	Using the spectral decomposition of the density matrix $ \rho_{Q}(t)=\sum_{n}^{}p_n\ket{n}\bra{n} $, where $ p_n(\ket{n}) $ are the eigenvalues (eigenvectors) of the density matrix $ \rho_{Q}(t) $, satisfying $ \sum_{n}^{}p_n=1 $ and $ p_n\geq0 $. The QFI reduces to the following form:
	\begin{equation}\label{QFI}
		\mathcal{F}(\rho_{Q}(t),H^{Q})=\dfrac{1}{2}\sum_{n\neq m}^{}\dfrac{(p_n-p_m)^2}{p_n+p_m}|\bra{n}H^{Q}\ket{m}|^2,
	\end{equation}
	the summation in the expression above is performed under the condition $ p_n+p_m>0 $. Effectively, we aim to achieve robust estimation precision, i.e., large quantum Fisher information, and vice versa. However, QFI is related to the concept of quantum entanglement (in general, non-classical correlations), which is the basis of superdense coding. Indeed, maximally entangled states allow to robust values of QFI \cite{Ref15}, which results in superior dsuperense coding capabilities. Recent results examined the relationship between entanglement and quantum metrology \cite{Ref21}. In fact, quantum entanglement has been shown to significantly enhance the accuracy of parameter estimation, as we will see in our results. 
	
	\subsection{Local quantum uncertainty}
	\label{sub:3}
	The local quantum uncertainty as a measure of discord, quantum correlations, was recently introduced by Girolami \emph{et al.} \cite{Girolami}. This measure captures non-classical information in a quantum state. However, while LQU is similar to quantum discord, it does not require a complex optimization procedure for the measurement. To quantify the quantum part, Wigner and Yanas proposed the concept of Skew information \cite{Ref24}
	\begin{equation}
		\mathcal{I}\big( \rho_{Q}(t),\mathbb{k}_{Q_1}\otimes \mathbb{1}_{Q_2} \big)=
		-\dfrac{1}{2}\hbox{Tr}\{ [\sqrt{\rho_{Q}(t)},\mathbb{k}_{Q_1}\otimes \mathbb{1}_{Q_2}] \},
	\end{equation}
	where $ \mathbb{k}_{Q_1} $ defines a local observable (Hermitian operator) acting on the qubit $ Q_1 $, and $ \mathbb{1}_{Q_2} $ denotes the identity operator acting on the qubit $ Q_2 $. Hence, the LQU with respect to the qubit $ Q_1 $ is defined as the minimum Skew information
	\begin{equation}
		\mathcal{Q}(\rho_{Q}(t))=\min_{\mathbb{k}_{Q_1}}\mathcal{I}\big( \rho_{Q}(t),\mathbb{k}_{Q_1}\otimes \mathbb{1}_{Q_2} \big).
	\end{equation}
	The Skew information under classical mixing is non-increasing and non-negative. The compact expression for LQU in $ 2\otimes d $ quantum state is expressed as follow \cite{Girolami}
	\begin{equation}
		\mathcal{Q}(\rho_{Q}(t))=1-\max\{ r_1,r_2,r_3 \},
	\end{equation}
	where $ r_i(i=1,2,3) $ are the eigenvalues of the ($3\times3$) symmetric matrix $ \mathcal{R} $. Indeed, the relation that provides the matrix elements of the symmetric matrix $ \mathcal{R} $ is defined as
	\begin{equation}\label{Symmetric matrix}
		\mathcal{R}_{ij}=\hbox{Tr}\{ \sqrt{\rho_{Q}(t)}\big( \sigma_{i}\otimes \mathbb{1}_{Q_2}\big)
		\sqrt{\rho_{Q}(t)}\big( \sigma_{j}\otimes \mathbb{1}_{Q_2} \big) \},
	\end{equation}
	where $ \sigma_{i}(i=x,y,z) $ are the Pauli matrices.

	\subsection{Trace distance measure}
	\label{sub:4}
	In Markovian processes, the distinction between any two quantum states gradually reduces over time due to the dynamics involved, resulting in a continuous loss of information to the surrounding environment. The presence of a temporal information exchange from the environment to the open system gives rise to quantum memory effects. This flow of information from the environment influences the dynamics of the open system, leading to the emergence of memory effects \cite{Ref25,Ref26}. However, to establish the so-called non-Markovianity measure, it is necessary to define a measurement that quantifies the distance between two quantum states represented, respectively, by two density matrices, namely $ \rho_{Q_1}^1 $ and $ \rho_{Q_1}^2 $. This measurement can be determined using the trace distance, which is defined as follows:
	\begin{equation}\label{eq:15}
		D\big( \rho_{Q_1}^1,\rho_{Q_1}^2 \big)=\dfrac{1}{2}\hbox{Tr}\{|\rho_{Q_1}^1-\rho_{Q_1}^2|\},
	\end{equation}
	where the norm of an operator $A$ is given by $ |A|=\sqrt{A^\dagger A} $.\\
	
	According to Eq.(\ref{eq:15}), the distinguishability of quantum states has a straightforward physical interpretation. Let Alice and Bob be two partners. She prepares a quantum system in one of two states: $ \rho_{Q_1}^1 $ or $ \rho_{Q_1}^2 $, each with a probability of $1/2$. The system is then given to Bob, who determines using a single measurement whether the system is in state $ \rho_{Q_1}^1 $ or $ \rho_{Q_1}^2 $ \cite{Found}. Then, the connection between the trace distance and the maximal success probability achievable by Bob through an optimal strategy can be demonstrated as
	\begin{equation}
		P_{max}	=\dfrac{1}{2}\big( 1+	D\big( \rho_{Q_1}^1,\rho_{Q_1}^2 \big) \big).
	\end{equation}
	If Alice prepared two orthogonal states, then $ D\big( \rho_{Q_1}^1,\rho_{Q_1}^2 \big)=1 $, which means that the maximal success probability achieved is unity, which indicates that Bob is able to distinguish the states with certainty. Another notable feature of the trace distance is the property that all completely positive and trace-preserving maps are contractions for this measure \cite{Found},
	\begin{equation}
		D\big( \rho_{Q_1}^1(t),\rho_{Q_1}^2(t) \big)\leq 	D\big( \rho_{Q_1}^1,\rho_{Q_1}^2 \big).
	\end{equation}
	This means that it is impossible for any completely positive and trace-preserving map to increase the distinguishability between two states. On the basis of these, one can express the rate of change of the trace distance of a pair of states as follows:
	\begin{equation}\label{DerTD}
		\sigma\big( t,\rho_{Q_1}^{1,2}(0) \big)=\dfrac{d}{dt} D\big( \rho_{Q_1}^1(t),\rho_{Q_1}^2(t) \big).
	\end{equation}
	
	\section{Suggested model}
	\label{sec:2}
\begin{figure}[h]
		\centering
		\resizebox{0.8\hsize}{!}{\includegraphics*{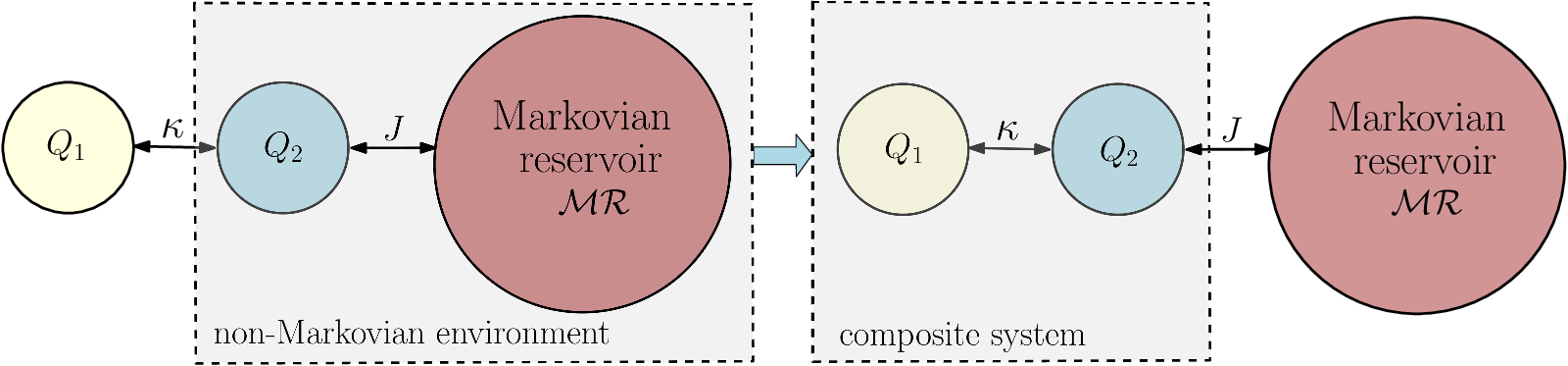}}
		\caption{Visual illustration of the suggested model.}
		\label{Fig.1}
	\end{figure}

	Let's consider a bipartite system composed of two coupled qubits, namely $ Q_1 $ and $ Q_2 $. This bipartite system is coupled to a Markovian reservoir, called $ \mathcal{MR} $. We assume that only the qubit $ Q_2 $ is essentially coupled to the reservoir $ \mathcal{MR} $. The model is composed of two main parts: the two-level qubit $ Q_1 $ and the structured non-Markovian environment (qubit $Q_2$ and the Markovian reservoir $\mathcal{MR}$) as shown in Fig.(\ref{Fig.1}).
	
	\noindent The Hamiltonian of the total system : the two qubits and the Markovian reservoir is in the following form:
	\begin{equation}\label{eq:1}
		H=H^{Q} +\omega^{\mathcal{MR}}\sigma_{+}^{\mathcal{MR}} \sigma_{-}^{\mathcal{MR}}+\frac{J}{2}\big( \sigma_{x}^{Q_2} \sigma_{x}^{\mathcal{MR}} + \sigma_{y}^{Q_2} \sigma_{y}^{\mathcal{MR}} \big),
	\end{equation}
	where $Q$ stands for the composite system $Q_1$ and $Q_2$, which in our case describes the open quantum system. Moreover, $ \omega^{\mathcal{MR}} $ is the frequency associated with the reservoir $ \mathcal{MR} $. Moreover, $ \sigma_{+} \big(\sigma_{-}\big) $ is the raising (lowering) operator, $ \sigma_{i}(i=x,y,z) $ are the Pauli matrices, and $ J $ is the coupling constant between $ Q_2 $ and $ \mathcal{MR} $. The first term in the right-hand side of the Eq.(\ref{eq:1}) is the Hamiltonian of the pair-qubits, the second term is the free Hamiltonian of the reservoir, which is assumed to be a qubit, and the last term is the Hamiltonian that describes the interaction between $ Q_2 $ and $ \mathcal{MR} $. The Hamiltonian of the pair-qubits has the following mathematical expression:
	\begin{equation}\label{eq:2}
		H^{Q}=\sum_{j=Q_1,Q_2}^{}
		\omega^{j}\sigma_{+}^{j} \sigma_{-}^{j}+\frac{\kappa}{2}
		\big( \sigma_{x}^{Q_1} \sigma_{x}^{Q_2} + 
		\sigma_{y}^{Q_1} \sigma_{y}^{Q_2} \big),
	\end{equation}
	where $ \omega^{Q_1}(\omega^{Q_2}) $ denotes the frequency of the qubit $ Q_1(Q_2) $, while the first term in the expression Eq.(\ref{eq:2}) describes the self Hamiltonian of each qubit. However, the second term describes the interaction between $ Q_1 $ and $ Q_2 $, through the coupling strength $ \kappa $. To solve the dynamics of this system, we shall solve the following Lindblad master equation \cite{Ref1.6}:
	\begin{eqnarray}\label{eq:3}
		\dot{\rho}_{Q}(t)&=&-i[H^{Q}, \rho_{Q}(t) ]
		+\gamma_{+}\big(\sigma_{+}^{Q_2} \rho_{Q}(t) \sigma_{-}^{Q_2}-\frac{1}{2}
		\big[\sigma_{-}^{Q_2}\sigma_{+}^{Q_2},\rho_{Q}(t)\big]_{+} \big)\nonumber\\
		&+&\gamma_{-}\big(\sigma_{-}^{Q_2} \rho_{Q}(t) \sigma_{+}^{Q_2}-\frac{1}{2}
		\big[\sigma_{+}^{Q_2}\sigma_{-}^{Q_2},\rho_{Q}(t)\big]_{+} \big),
	\end{eqnarray}
	where  $\gamma_{+}=\gamma\braket{\sigma_{+}^{\mathcal{MR}}\sigma_{-}^{\mathcal{MR}}}$ and $\gamma_{-}=\gamma\braket{\sigma_{-}^{\mathcal{MR}}\sigma_{+}^{\mathcal{MR}}}$ are the decay rates resulting from the interaction of the system with the reservoir. Generally, $ \big[A,B\big]_{+} $ defines the anti-commutator of $A$ and $B$. The expectation values in the decay rates are computed by performing a simple calculation and using the initial state of the reservoir as a thermal state, namely $ \rho_{\mathcal{MR}}=e^{-\beta^{\mathcal{MR}} \omega^{\mathcal{MR}} H^{\mathcal{MR}}} $. In fact, one can get,
	\begin{eqnarray}\label{eq:4}
		\braket{\sigma_{+}^{\mathcal{MR}}\sigma_{-}^{\mathcal{MR}}}_{\rho_{\mathcal{MR}}}=\frac{1}{1+e^{\beta^{\mathcal{MR}} \omega^{\mathcal{MR}}}}\;\;\;\;,\;\;\;\;
		\braket{\sigma_{-}^{\mathcal{MR}}\sigma_{+}^{\mathcal{MR}}}_{\rho_{\mathcal{MR}}}=
		\frac{1}{1+e^{-\beta^{\mathcal{MR}} \omega^{\mathcal{MR}}}}.
	\end{eqnarray}
	
	The analytical results of the above preliminaries using the suggested model in Sec. \ref{sec:2} are given in \textbf{Appendix A}. Indeed, it is assumed that the pair-qubits are initially prepared in the state,
		\begin{equation}\label{initial}
			\ket{\psi(0)}_{Q}=\cos(\frac{\theta}{2})\ket{00}+e^{i\varphi}\sin(\frac{\theta}{2})\ket{11}, 
		\end{equation}
		where $0\leq\theta\leq\pi$ and  $0\leq\phi\leq 2 \pi$. Finally, we note that by using Eqs.(\ref{eq:3}) and (\ref{initial}), the reduced density matrix of the pair-qubits is obtained in a special form called X-state \cite{X state} (the elements are given in \textbf{Appendix A}). In addition, in the rest of the paper, we set $ \omega^{Q_1}=\omega^{Q_2}=\omega^{\mathcal{MR}}=\omega$.\\

\begin{figure}[h]
		\begin{subfigure}{.33\textwidth}
			\centering
			\subcaption[short for lof]{}
			\includegraphics[width=.9\linewidth]{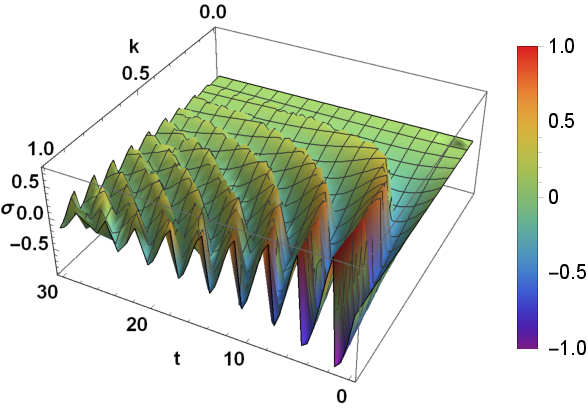}  
			\label{a2}
		\end{subfigure}		
		\begin{subfigure}{.33\textwidth}
			\centering
			\subcaption[short for lof]{}
			\includegraphics[width=.9\linewidth]{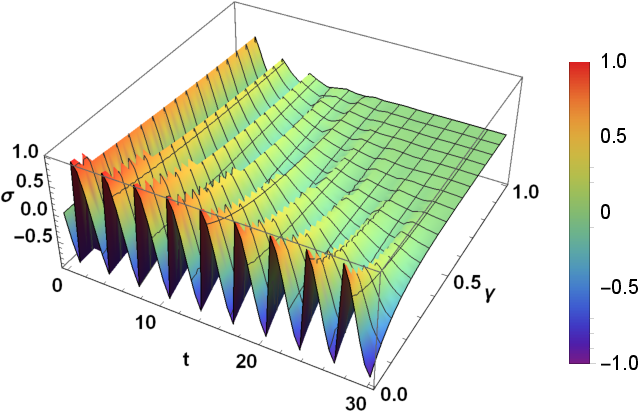}  
			\label{b2}
		\end{subfigure}
		\begin{subfigure}{.33\textwidth}
			\centering
			\subcaption[short for lof]{}
			\includegraphics[width=.9\linewidth]{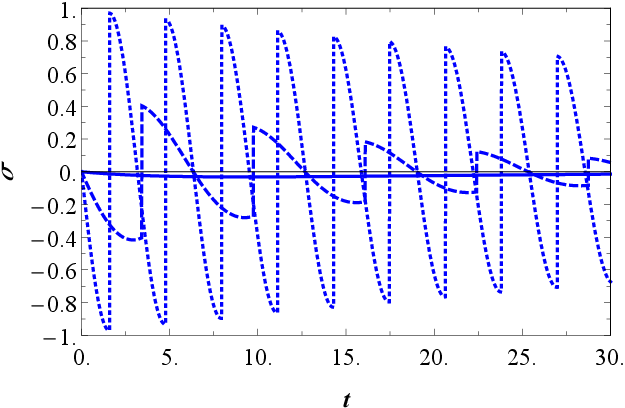}  
			\label{c2}
		\end{subfigure}
		\caption{Evolution of $\sigma\big( t,\rho_{Q_1}^{1,2}(0) \big) $ (a) versus $ \kappa $ and $t$, for $ \gamma=0.3$. (b)  versus $ \gamma $ and $t$, for fixed $\kappa=0.99$. We set, $\omega=2.2, T=0.1 pK$. Moreover, we plot in (c) $\sigma\big( t,\rho_{Q_1}^{1,2}(0) \big) $ versus $t$, for $\dfrac{\kappa}{\gamma}= 20,2$ and $0.2$ for dotted, dashed and solid lines, respectively.} 
		
		\label{fig:2}
	\end{figure}

	In this work, the time derivative of the trace distance defined in Eq.(\ref{DerTD}) will be used to indicate the non-Markovian regime. Indeed, a process is considered non-Markovian if and only if the derivative of trace distance with respect to time is positive at certain intervals of time. Physically, this means that information is flowing back from the environment into the open system. In Fig.\ref{fig:2}, we show that considering robust values of the coupling strength $\kappa$, the non-Markovian dynamic is guaranteed. \\
	
	In Fig.(\ref{fig:2}), the impact of the coupling strength between the pair-qubits, namely $\kappa$ and the decoherence parameter $\gamma$ on the derivative of trace distance is examined. The proposed initial states for the qubit $Q_1$ are $ \rho_{Q_1}^{1}(0)=\ket{+}\bra{+} $ and $ \rho_{Q_1}^{2}(0)=\ket{-}\bra{-} $, while the qubit $ Q_2 $ is initially prepared in the state $\rho_{Q_2}^{1,2}(0)=\ket{0}\bra{0}$. In particular, Fig.(\ref{a2}) evaluates the appropriate values of $\kappa$ that allow for a non-Markovian regime. It is clear that as $\kappa$ decreases, the non-Markovian behavior diminishes until it disappears. This is equivalent to saying that strong coupling between the pair-qubits gives rise to non-Markovian dynamics. In Fig.(\ref{b2}) the behavior of $\sigma\big( t,\rho_{Q_1}^{1,2}(0) \big)$ vanishes by increasing the decoherence parameter. While, when the parameter $\gamma$ is near zero, the coupling between the qubit $ Q_1 $ and the reservoir $ \mathcal{MR} $ is negligible, i.e., there are no decoherence effects, which means that there is a conservation of information between the qubit $ Q_1 $ and $ Q_2 $.\\
	
	In addition, we also show simultaneously the impact of $\kappa$ and $\gamma$ on the evolution of the derivative $\sigma\big( t,\rho_{Q_1}^{1,2}(0) \big)$ in Fig.(\ref{c2}). Clearly, the plot shows that by imposing $\kappa/\gamma=20$ and $\kappa/\gamma=2$, the derivative of the trace distance turns positive and thus the memory effects between the open system and its environment are highly apparent, i.e., non-Markovian regime. However, by choosing $\kappa/\gamma=0.2$ (solid line), the derivative of trace distance is close to zero, which indicates that the dynamics are Markovian.
	
	
	\section{Exact solutions and discussion}
	\label{sec:4}
	In this section, we investigate the effect of the parameters $\kappa, \omega,\gamma, T$ encoded in the reduced density operator $\rho_{Q}(t)$ and the weight parameter $\theta$ encoded in the initial state on the behavior of SDC, QFI, LQU and $\sigma\big( t,\rho_{Q_1}^{1,2}(0) \big)$. We aim to show whether these parameters can enhance the memory effects resulting from system-environment interaction. 
	\subsection{Effect of the weight parameter $\theta$}

	\begin{figure}[h]
		\begin{subfigure}{.5\textwidth}
			\centering
			\subcaption[short for lof]{}
			\includegraphics[width=0.8\linewidth]{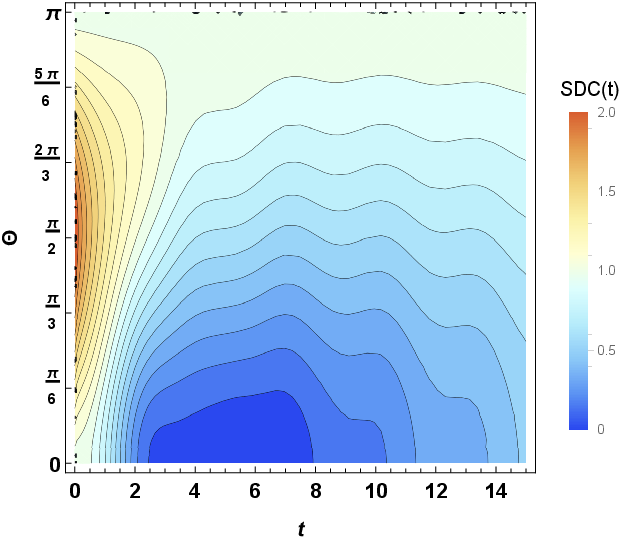}  
			\label{a4}
		\end{subfigure}
		\begin{subfigure}{.5\textwidth}
			\centering
			\subcaption[short for lof]{}
			\includegraphics[width=0.8\linewidth]{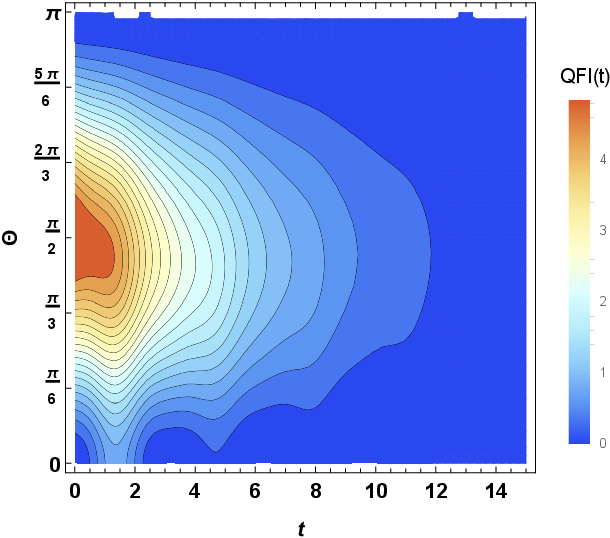}  
			\label{b4}
		\end{subfigure}
		\begin{subfigure}{.5\textwidth}
			\centering
			\subcaption[short for lof]{}
			\includegraphics[width=0.8\linewidth]{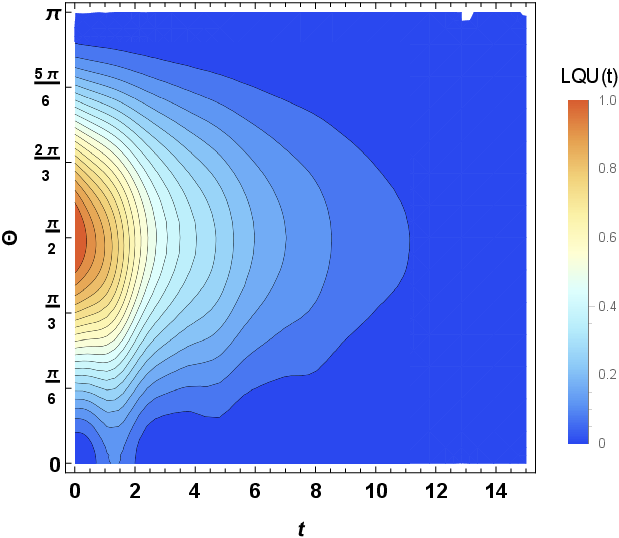}  
			\label{c4}
		\end{subfigure}
		\begin{subfigure}{.5\textwidth}
			\centering
			\subcaption[short for lof]{}
			\includegraphics[width=0.84\linewidth]{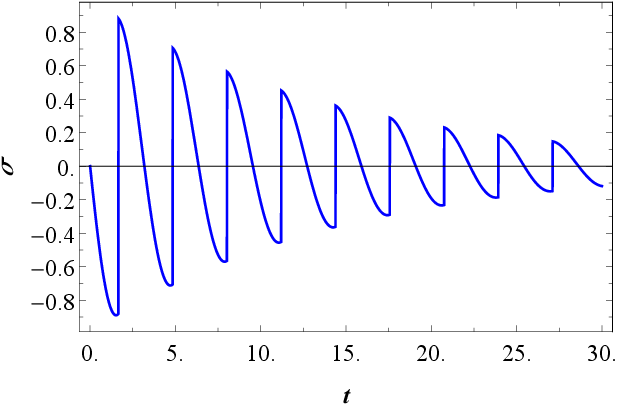}  
			\label{d4}
		\end{subfigure}	
		\caption{Contour plots of (a) SDC, (b) QFI, (c) LQU as a function of $t$ and $ \theta $  and (d) illustrate $ \sigma\big( t,\rho_{Q_1}^{1,2}(0) \big) $ versus $t$. Moreover, we set $ \omega=2.2,\; \kappa=0.99,\; T=0.1 pK,\; \gamma=0.3$.}
		\label{fig:4}
	\end{figure}
	
	In fact, Fig.(\ref{fig:4}) reflects the non-Markovian regime, i.e., the coupling between $ Q_1 $ and $ Q_2 $ is strong ($\kappa=0.99$). Figs.(\ref{a4}), (\ref{b4}),(\ref{c4}) exhibit the contour plots of superdense coding, QFI, and LQU versus time interaction $ t $ and $ \theta $, while Fig.(\ref{a4}) shows that the behavior of superdense coding decreases when time interaction $t$ increases. Interestingly enough, the superdense coding initially reaches its maximum value, which is 2 for $ \theta=\pi/2 $ (This can be verified by setting '$ t $' equals to  0 in Eq.(\ref{SDC's formula})). This means that superdense coding achieves the optimal value only initially for the maximally entangled state, but collapses asymptotically due to decoherence. Obviously, for a specific interval of interaction time, $t\in [1,3]$, the superdense coding becomes valid, but not optimal. However, when $ \theta $ either decreases in the interval $ [0, \pi/2] $ or increases in the interval $ [\pi, \pi/2] $, the initial values of the superdense coding decrease; the latter is valid ($1<SDC<2$) but never optimal, $SDC <2$. Fig. (\ref{b4}) presents the dynamics of QFI for the same parameters as for superdense coding in Fig.(\ref{a4}). We clearly observe that the amount of QFI encoded in the density matrix $\rho_Q(t)$ is also initially maximized for $ \theta=\pi/2 $ (See Eq.(\ref{QFI's formula})), which means that the state is initially prepared in a maximally entangled state, as it is clear from Eq.(\ref{initial}). For $t>0$, QFI reduces by decreasing $ \theta $ in the region $ [0, \pi/2] $ or increasing it in the region $ [\pi,\pi/2] $, and vice versa. \\
	
	For the sake of comparison,  we plot in Fig.(\ref{c4}) the behavior of LQU as a function of time and $ \theta $. From the results shown in this figure and from  Eq.(\ref{LQU's formula}), it is clear that the maximum amount of LQU is also obtained for $ \theta=\pi/2 $ as in superdense coding and QFI. One can observe a great similarity between the behavior of LQU and that of QFI. Therefore, we can conclude that strong quantum correlations enhance the sensitivity of the reduced density operator $\rho_Q(t)$ by means of QFI. This means that strong correlations encoded in $\rho_Q(t)$ improve the precision of a measurement of a quantum state, which gives rise to the delicate relationship between quantum metrology (QFI) and quantum information theory (superdense coding, LQU). \\
	
	In addition, we can conclude that LQU can be used as a tool to detect optimal superdense coding capacity and robust sensitivity in the reduced state. However, these results are described in the non-Markovian regime corresponding to the robust coupling strength, where we set $\kappa=0.99$. However, the behavior of SDC, QFI, LQU and $\sigma\big( t,\rho_{Q_1}^{1,2}(0)$ shows many oscillations over $t$, which means that the coherence of the composite system increases during certain times, which could be interpreted as the backflow of information from the reservoir $ \mathcal{MR} $ to the composite system $ Q $ due to the memory effects between them. In fact, Fig.(\ref{d4}) demonstrates that $\sigma\big( t,\rho_{Q_1}^{1,2}(0)$ is positive for some time periods. In other words, these oscillations can be explained by the strong coupling between $Q_1$ and $Q_2$. This robust coupling led to the emergence of non-Markovianity dynamics during interaction with the reservoir $ \mathcal{MR} $. On the basis of this, we can conclude that the qubit $ Q_2 $ plays a crucial role in controlling the non-Markovian behavior of the proposed system.
	
	\subsection{Effect of the temperature $T$}

	\begin{figure}[h!]
		\begin{subfigure}{.5\textwidth}
			\centering
			\subcaption[short for lof]{}
			\includegraphics[width=0.8\linewidth]{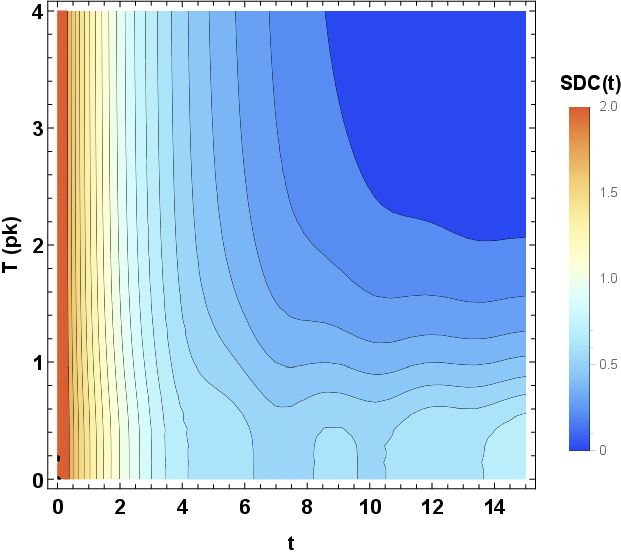}  
			\label{a6}
		\end{subfigure}
		\begin{subfigure}{.5\textwidth}
			\centering
			\subcaption[short for lof]{}
			\includegraphics[width=0.8\linewidth]{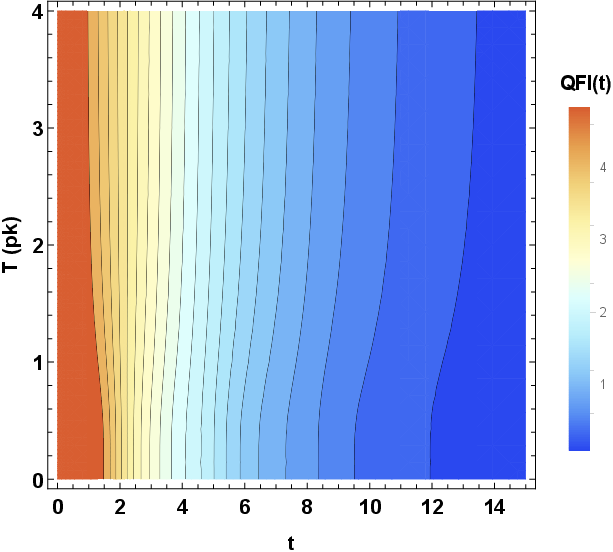}  
			\label{b6}
		\end{subfigure}
		\begin{subfigure}{.5\textwidth}
			\centering
			\subcaption[short for lof]{}
			\includegraphics[width=0.8\linewidth]{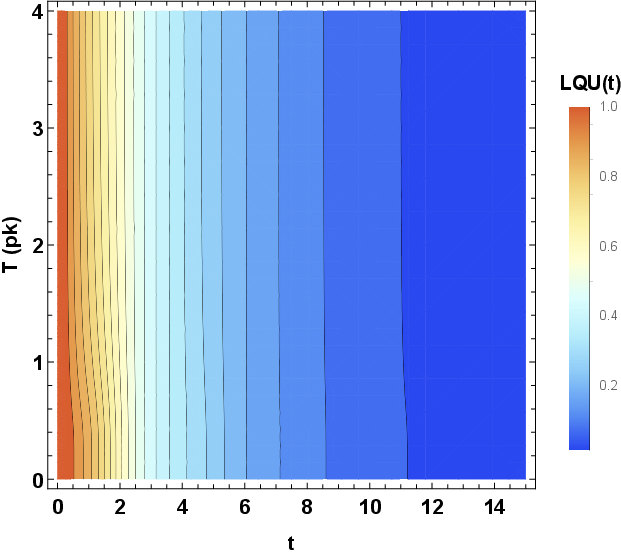}  
			\label{c6}
		\end{subfigure}
		\begin{subfigure}{.5\textwidth}
			\centering
			\subcaption[short for lof]{}
			\includegraphics[width=0.84\linewidth]{dTDNM1.eps}  
			\label{d6}
		\end{subfigure}	
		\caption{Contour plots of (a) SDC, (b) QFI, (c) LQU as a function of $t$ and $ T $  and (d) illustrate $ \sigma\big( t,\rho_{Q_1}^{1,2}(0) \big) $ versus $t$. Moreover, we set $ \omega=2.2,\; \kappa=0.99,\; \theta=\dfrac{\pi}{2},\; \gamma=0.3$.}
		\label{fig:6}
	\end{figure}

	Fig.(\ref{fig:6}) illustrates the contour plots against the time interaction parameter $ t $ and the temperature $T$ in the non-Markovian regime. It is clear that the behavior of superdense coding in a specific interval of temperature is the same: $T\in [0.01pK,0.2pK] $. In this case, it is obvious that superdense coding is initially optimal $SDC=2$ as it is also clear from Eq.(\ref{SDC's formula}). Once the dynamics is functional, i.e., $t>0$, the superdense coding gradually decreases. For large values of the temperature parameter, the superdense coding is reduced, as in the previous case, but it is smaller than that obtained for smaller temperatures i.e., large values of temperature in the long time interaction results in mixing the density matrix of the composite system $\rho_Q$(t). Figs.(\ref{b6}) and (\ref{c6}) show the contour plots of QFI and LQU as functions of the time interaction parameter $t$ and temperature $ T $, respectively. Both quantities are initially maximized and decrease asymptotically until they disappear when $t$ takes robust values. Again, they show similar behavior, which means that they provide the same information. Indeed, strong correlations allow for high sensitivity of the reduced density operator and vice versa.  Indeed, Fig.(\ref{d6}) affirms for $\kappa=0.99$ that the dynamics of the system is non-Markovian (positivity of $ \sigma\big( t,\rho_{Q_1}^{1,2}(0) \big)$). Hence, we notice that the non-Markovian behavior is sensitive to changing temperature $T$, where large values of the temperature destroy the non-Markovianity, QFI and LQU.\\
	
	\subsection{Effect of the decoherence $\gamma$}
	
	\begin{figure}[h]
		\begin{subfigure}{.5\textwidth}
			\centering
			\subcaption[short for lof]{}
			\includegraphics[width=0.8\linewidth]{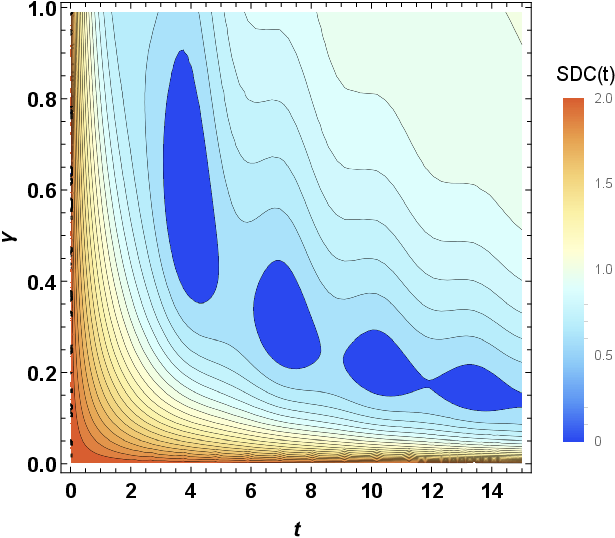}  
			\label{a8}
		\end{subfigure}
		\begin{subfigure}{.5\textwidth}
			\centering
			\subcaption[short for lof]{}
			\includegraphics[width=0.8\linewidth]{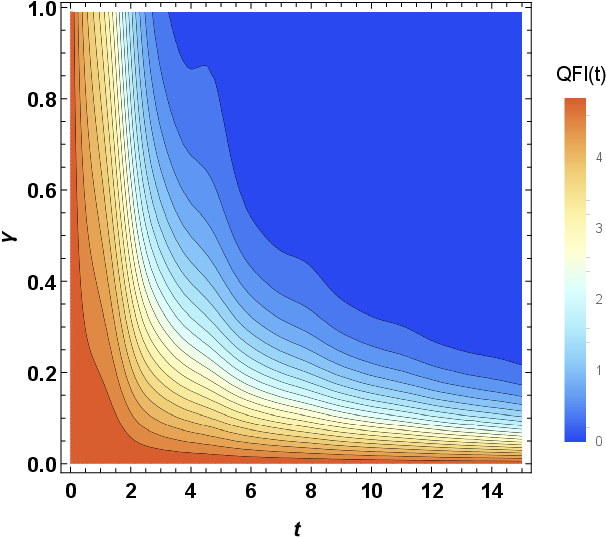}  
			\label{b8}
		\end{subfigure}
		\begin{subfigure}{.5\textwidth}
			\centering
			\subcaption[short for lof]{}
			\includegraphics[width=0.8\linewidth]{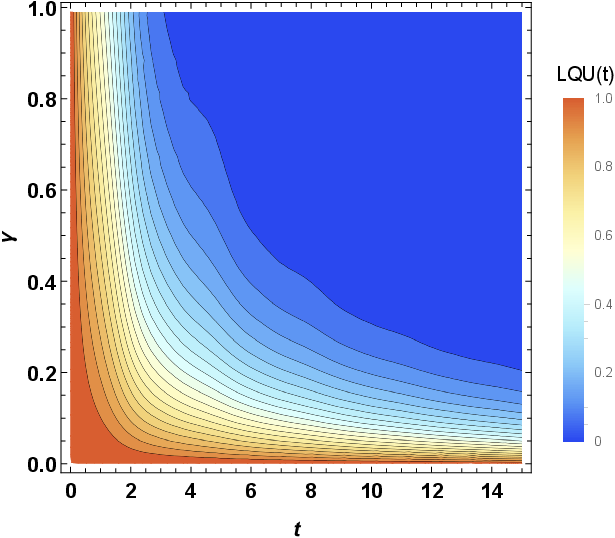}  
			\label{c8}
		\end{subfigure}
		\begin{subfigure}{.5\textwidth}
			\centering
			\subcaption[short for lof]{}
			\includegraphics[width=0.84\linewidth]{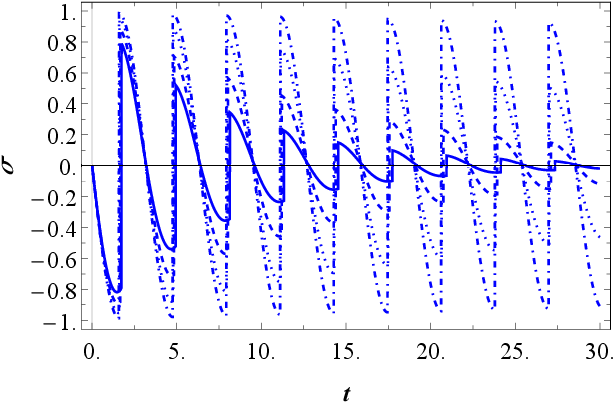}  
			\label{d8}
		\end{subfigure}	
		\caption{Contour plots of (a) SDC, (b) QFI, (c) LQU as a function of $t$ and $ \gamma $  and (d) illustrate $ \sigma\big( t,\rho_{Q_1}^{1,2}(0) \big) $ versus $t$ where ($ \gamma =0.5$  (Solid line), $ \gamma =0.3$ (Dashed line), $ \gamma =0.1$ (Dotted line),  $ \gamma =0.01$ (Dot-dashed line)). For fixed $ \omega=2.2,\; \kappa=0.99,\; T=0.1 pK,\; \theta=\dfrac{\pi}{2}$.}	
		\label{fig:8}
	\end{figure}

	The next step is to intuitively describe the influence of the decoherence parameter on the dynamics of different measures. From Fig.(\ref{a8}), we clearly see that the superdense coding is initially optimal $SDC=2$ which is satisfied exactly in Eq.(\ref{SDC's formula}), but as $t$ increases, it decreases asymptotically. For small values of the decoherence rate $\gamma$, it remains valid in a long-time interval, $1<SDC<2$, contrariwise, by increasing $\gamma$ the validity of superdense coding disappears. The behavior of QFI (Fig.(\ref{b8})) and LQU (Fig.(\ref{c8})) is similar to that of superdense coding with some oscillations, that is, the memory effects are increasing. In fact, Fig.(\ref{d8}) shows that $\sigma\big( t,\rho_{Q_1}^{1,2}(0)$ is positive at certain values of time. These oscillations are very large for small numbers of $\gamma$, while they gradually reduce when $\gamma$ increases. This can be interpreted as follows: lower values of $\gamma$ lead to the emergence of memory effects between the open system and its surrounding environment. Moreover, the similarity between these quantities indicates that they provide the same information about the system-environment interaction. \\

	From Fig.(\ref{fig:2})-(\ref{fig:8}), the qubit $ Q_2$ plays a significant role in generating and controlling the non-Markovianity, non-classical correlations, superdense coding and sensitivity of the state by means of QFI. They explain that by increasing the ratio $ \kappa/\gamma $, the memory effect arises, and vice versa. However, the initial values of the parameter $ \theta $ have a crucial influence on the behavior of SDC, QFI, and LQU. By increasing the decoherence parameter, the behavior of these measures quickly disappears. The same holds for the effect of temperature, where the memory effects increase by decreasing the temperature. In particular, when LQU takes its maximum value, i.e., $ LQU=1$, SDC becomes optimal and QFI is maximized. Therefore, several factors may protect and enhance the superdense coding protocol, namely lower temperatures leading to a stable behavior of SDC, and high non-classical correlations in terms of LQU resulting in optimal superdense coding. Furthermore, long-lived superdense coding can be achieved by decreasing the decoherence parameters.
	
	\section{Conclusion}
	\label{sec:5}
	In this paper, we investigated the relationship between superdense coding and non-classical correlations by means of local quantum uncertainty and quantum Fisher information. We evaluated the dynamics of these quantities under the non-Markovian regime of an open system that interacts with its surrounding environment. To discuss these phenomena, we assumed that two qubits are coupled to each other through a coupling constant of $\kappa$. We supposed that the qubits are initially prepared in a partially/entangled state. The two qubits interact first with a Markovian reservoir, namely $ \mathcal{MR} $, which is coupled only to the qubit $ Q_2 $ via another coupling strength $J$. The effect of different initial values of the weight parameter $ \theta $ encoded in the initial state, temperature, and the decoherence parameter $ \gamma $ are discussed in the non-Markovian regime. From the obtained results, we concluded that it is possible to attain a non-Markovian regime by controlling the strength coupling between $ Q_1 $ and $ Q_2 $. When the strength of coupling between the pair-qubits is stronger, we notice that oscillations appear in the behavior of SDC, QFI, and LQU. We interpreted those oscillations as the emergence of memory effects between the open system $Q$ and its reservoir. In other words, there is a backflow of information from the reservoir into the qubit $ Q_1 $ via $ Q_2 $, as demonstrated in the behavior of the derivative of trace distance.\\
	
	The general behavior of the dynamics of SDC, QFI, and LQU is strongly influenced by the initial values of the parameter $ \theta $. In fact, when $ \theta=\pi/2 $, the amount of LQU is maximized, i.e., strong non-classical correlations. In this case, superdense coding becomes optimal. Therefore, optimal superdense coding can be detected through the behavior of the LQU. On the other hand, we showed that the amount of QFI followed exactly the behavior of quantum correlations, which shed light on the delicate relationship between quantum metrology and quantum information theory. Our results showed that non-classical correlations enhance the accuracy of a measurement in a quantum state. In addition, the numerical results proved that QFI detects non-Markovian behavior. Furthermore, the osculations that appeared in the behaviors of SDC, LQU and QFI indicating the exchange of information between the open system $Q$ and its reservoir, are affirmed by the positive values of the derivative of trace distance.\\

	Finally, we conclude that these results lead to better controllability for generating non-Markovianity through the coupling strength between a pair of qubits. This work can be extended in several ways. For example, one can explore different initial states for the pair-states or compare local quantum uncertainty and negativity in other types of systems.
	
	\renewcommand{\theequation}{A-\arabic{equation}}
	\setcounter{equation}{0}  
	\section*{Appendix A: ELEMENTS OF THE REDUCED DENSITY MATRIX}
	
	In this appendix, we provide the analytical (exact) expressions for the elements of the density matrix $\rho_{Q}(t)$, which are derived straightforwardly from Eq. (\ref{eq:3}). This density matrix has a clear X-shaped structure as:
	
	\begin{eqnarray}\label{Densitu Matrix}
		\rho_{Q}(t)&=&\rho_{11}(t)\ket{00}\bra{00}+\rho_{14}(t)\ket{11}\bra{00}+\rho_{22}(t)\ket{01}\bra{01}+\rho_{23}(t)\ket{10}\bra{01}\nonumber\\
		&+&\rho_{32}(t)\ket{01}\bra{10}+\rho_{33}(t)\ket{10}\bra{10}+\rho_{41}(t)\ket{00}\bra{11}+\rho_{44}(t)\ket{11}\bra{11},
	\end{eqnarray}
where the different elements are: 
	
	\begin{eqnarray} 
		\rho_{11}(t)&=&\frac{\cos (\Theta ) e^{\gamma  (-t)} \left(\gamma ^2 e^{\frac{\gamma  t}{2}} \cosh \left(\frac{1}{2} t \sqrt{\gamma ^2-16 k^2}\right)-8 k^2 \left(2 e^{\frac{\gamma  t}{2}}+e^{\omega /T}-1\right)\right)}{\left(\gamma ^2-16 k^2\right) \left(e^{\omega /T}+1\right)}\nonumber\\
		&+&\frac{e^{-\frac{1}{2} (\gamma  t)} \left(\gamma ^2 \left(e^{\omega /T}-1\right) \cosh \left(\frac{1}{2} t \sqrt{\gamma ^2-16 k^2}\right)-\left(16 k^2-\gamma ^2\right) \left(e^{\frac{2 \omega }{T}}+1\right) \cosh \left(\frac{\gamma  t}{2}\right)\right)}{\left(\gamma ^2-16 k^2\right) \left(e^{\omega /T}+1\right)^2}\nonumber\\
		&+&\frac{e^{\gamma  (-t)} \left(e^{\omega /T}-1\right) \left(\gamma ^2 \cos (\Theta ) \left(e^{\omega /T}+1\right)-32 k^2 e^{\frac{\gamma  t}{2}}\right)}{2 \left(\gamma ^2-16 k^2\right) \left(e^{\omega /T}+1\right)^2}-\frac{1}{2} \tanh \left(\frac{\omega }{2 T}\right)\nonumber\\
		\rho_{22}(t)&=&\frac{e^{\gamma  (-t)} \left(\left(\gamma ^2 \left(e^{\gamma  t}+1\right)-16 k^2 \left(e^{\frac{\gamma  t}{2}}-1\right)^2\right) \text{sech}^2\left(\frac{\omega }{2 T}\right)-2 \left(\gamma ^2+16 k^2 \left(e^{\frac{\gamma  t}{2}}-1\right)\right)\right)}{4 \left(\gamma ^2-16 k^2\right)}\nonumber\\
		&+& \frac{\gamma  e^{-\frac{1}{2} (\gamma  t)} \text{sech}^2\left(\frac{\omega }{2 T}\right) \sinh \left(\frac{1}{2} t \sqrt{\gamma ^2-16 k^2}\right) \left(\cos (\Theta )+\cos (\Theta ) \cosh \left(\frac{\omega }{T}\right)+\sinh \left(\frac{\omega }{T}\right)\right)}{4 \sqrt{\gamma ^2-16 k^2}}\nonumber\\
		&+& \frac{\cos (\Theta ) e^{\gamma  (-t)} \tanh \left(\frac{\omega }{2 T}\right) \left(-\gamma ^2+e^{\frac{\gamma  t}{2}} \left(\gamma ^2 \cosh \left(\frac{1}{2} t \sqrt{\gamma ^2-16 k^2}\right)-16 k^2\right)+16 k^2\right)}{2 \left(\gamma ^2-16 k^2\right)}\nonumber\\
		&+&\frac{\gamma ^2 e^{-\frac{1}{2} (\gamma  t)} \tanh ^2\left(\frac{\omega }{2 T}\right) \cosh \left(\frac{1}{2} t \sqrt{\gamma ^2-16 k^2}\right)}{2 \gamma ^2-32 k^2}\nonumber\\
		\rho_{33}(t)&=&-\frac{4 k^2 e^{\gamma  (-t)} \left(e^{\frac{\gamma  t}{2}}-1\right) \text{sech}^2\left(\frac{\omega }{2 T}\right) \left(e^{\frac{\gamma  t}{2}}+\cos (\Theta ) \sinh \left(\frac{\omega }{T}\right)+\cosh \left(\frac{\omega }{T}\right)\right)}{\gamma ^2-16 k^2}\nonumber\\	
		&+&\frac{\gamma ^2 e^{\gamma  (-t)} \text{sech}^2\left(\frac{\omega }{2 T}\right) \left(-e^{\frac{\gamma  t}{2}} \cosh \left(\frac{1}{2} t \sqrt{\gamma ^2-16 k^2}\right)+e^{\gamma  t}-\cos (\Theta ) \sinh \left(\frac{\omega }{T}\right)-\cosh \left(\frac{\omega }{T}\right)\right)}{4 \left(\gamma ^2-16 k^2\right)}\nonumber\\
		&+&\frac{\gamma ^2 e^{-\frac{1}{2} (\gamma  t)} \text{sech}^2\left(\frac{\omega }{2 T}\right) \cosh \left(\frac{1}{2} t \sqrt{\gamma ^2-16 k^2}\right) \left(\cos (\Theta ) \sinh \left(\frac{\omega }{T}\right)+\cosh \left(\frac{\omega }{T}\right)\right)}{4 \left(\gamma ^2-16 k^2\right)}\nonumber\\
		&-&\frac{\gamma  e^{-\frac{1}{2} (\gamma  t)} \text{sech}^2\left(\frac{\omega }{2 T}\right) \sinh \left(\frac{1}{2} t \sqrt{\gamma ^2-16 k^2}\right) \left(\cos (\Theta )+\cos (\Theta ) \cosh \left(\frac{\omega }{T}\right)+\sinh \left(\frac{\omega }{T}\right)\right)}{4 \sqrt{\gamma ^2-16 k^2}}\nonumber\\
		\rho_{44}(t)&=&1-\big(\rho_{11}(t)+\rho_{22}(t)+\rho_{33}(t)\big),\nonumber\\
		\rho_{23}(t)&=& \frac{4 i \gamma  k e^{-\frac{1}{2} (\gamma  t)} \sinh ^2\left(\frac{1}{4} t \sqrt{\gamma ^2-16 k^2}\right) \left(\cos (\theta )+\tanh \left(\frac{\omega }{2 T}\right)\right)}{\gamma ^2-16 k^2}=\rho_{32}^{*}(t),\nonumber\\
		\rho_{14}(t)&=& \frac{1}{2} \sin (\theta ) e^{-\frac{\gamma  t}{2}+i (2 t \omega - \varphi) }=\rho_{41}^{*}(t).
	\end{eqnarray}	
	
	The eigenvalues of the density matrix $\rho_Q(t)$ are:
	\begin{eqnarray}\label{eige}
		p_{1,2}(t)=\frac{1}{2} \left(\rho _{11}(t)+\rho _{44}(t)\pm\sqrt{\left(\rho _{11}(t)-\rho _{44}(t)\right){}^2+4 \rho _{14}(t) \rho _{41}(t)}\right),\nonumber\\
		p_{3,4}(t)=\frac{1}{2} \left(\rho _{22}(t)+\rho _{33}(t)\pm\sqrt{\left(\rho _{22}(t)-\rho _{33}(t)\right){}^2+4 \rho _{23}(t) \rho _{32}(t)}\right).
	\end{eqnarray}
	The corresponding eigenvectors are
	\begin{eqnarray}\label{eigenv}
		\ket{\phi_1}=\big(a_1(t),0,0,b_1(t)\big)^t\nonumber\\
		\ket{\phi_2}=\big(a_2(t),0,0,b_2(t)\big)^t\nonumber\\
		\ket{\phi_3}=\big(0,a_3(t),b_3(t),0\big)^t\nonumber\\
		\ket{\phi_4}=\big(0,a_4(t),b_4(t),0\big)^t
	\end{eqnarray}
	where
	\begin{eqnarray}
		a_1(t)&=&\dfrac{p_1-\rho_{44}(t)}{\rho_{41}(t) \sqrt{\bigg| \dfrac{p_1-\rho _{44}(t)}{\rho_{41}(t)}\bigg|^2+1}},\;\;\;\;
		b_1(t)=\dfrac{1}{\sqrt{\bigg| \dfrac{p_1-\rho _{44}(t)}{\rho_{41}(t)}\bigg|^2+1}}\nonumber\\
		a_2(t)&=&\dfrac{p_2-\rho_{44}(t)}{\rho_{41}(t) \sqrt{\bigg| \dfrac{p_2-\rho _{44}(t)}{\rho_{41}(t)}\bigg|^2+1}},\;\;\;\;
		b_2(t)=\dfrac{1}{\sqrt{\bigg| \dfrac{p_2-\rho _{44}(t)}{\rho_{41}(t)}\bigg|^2+1}}\nonumber\\
		a_3(t)&=&\dfrac{p_3-\rho_{33}(t)}{\rho_{32}(t) \sqrt{\bigg| \dfrac{p_3-\rho _{33}(t)}{\rho_{32}(t)}\bigg|^2+1}},\;\;\;\;
		b_3(t)=\dfrac{1}{\sqrt{\bigg| \dfrac{p_3-\rho _{33}(t)}{\rho_{32}(t)}\bigg|^2+1}}\nonumber\\
		a_4(t)&=&\dfrac{p_4-\rho_{33}(t)}{\rho_{32}(t) \sqrt{\bigg| \dfrac{p_4-\rho _{33}(t)}{\rho_{32}(t)}\bigg|^2+1}},\;\;\;\;
		b_4(t)=\dfrac{1}{\sqrt{\bigg| \dfrac{p_4-\rho _{33}(t)}{\rho_{32}(t)}\bigg|^2+1}}
	\end{eqnarray}
	
	\renewcommand{\theequation}{B-\arabic{equation}}
	\setcounter{equation}{0}  
	\section*{Appendix B: DERIVATION THE EXPRESSION OF OF SDC, QFI AND LQU}
	
	In this appendix, we provide the exact expressions for the used measures, namely SDC, QFI and LQU, introduced in Sec.\ref{sec:3}.\\
	
	$\bullet$ \textbf{Superdense coding capacity}\\
	
	By applying the mutually orthogonal unitary transformation and by inserting $ \rho_{Q}(t) $ from Eq.(\ref{Densitu Matrix}) into Eq.(\ref{Average}), the average state of the signal ensemble is derived as: 
	\begin{eqnarray}
		\overline{\rho_{Q}^*}(t)&=&\frac{1}{2} \left(\rho _{22}(t)(t)+\rho _{33}(t)\right)\ket{00}\bra{00}+\frac{1}{2} \left(\rho _{22}(t)+\rho _{44}(t)\right)\ket{01}\bra{01}\nonumber\\
		&+&\frac{1}{2} \left(\rho _{22}(t)(t)+\rho _{33}(t)\right)\ket{10}\bra{10}+\frac{1}{2} \left(\rho _{22}(t)+\rho _{44}(t)\right)\ket{11}\bra{11}.
	\end{eqnarray}
	Hence, the superdense coding capacity takes the following form: 
	\begin{eqnarray}\label{SDC's formula}
		\mathcal{SDC}(t)&=&\bigg(  -\left(\rho _{22}(t)+\rho _{33}(t)\right)\log_2(\frac{1}{2} \left(\rho _{22}(t)+\rho _{33}(t)\right))
		-\left(\rho _{22}(t)+\rho _{44}(t)\right)\log_2(\frac{1}{2} \left(\rho _{22}(t)+\rho _{44}(t)\right)) \bigg)\nonumber\\
		&-&\bigg( -p_{1}(t)\log_2(p_{1}(t))-p_{2}(t)\log_2(p_{2}(t))
		-p_{3}(t)\log_2(p_{3}(t))-p_{4}(t)\log_2(p_{4}(t)) \bigg).
	\end{eqnarray}
	
	$\bullet$ \textbf{Quantum Fisher information}\\
	
	By employing expression of eigenvalues and eigenvectors already calculated in Eqs.(\ref{eige}) and (\ref{eigenv}) into Eq.(\ref{QFI}), we get the formula of quantum Fisher information as follows: 
	
	\begin{eqnarray}\label{QFI's formula}
		\mathcal{F}(\rho_{Q}(t),H^{Q})=\frac{({p_3}-{p_4})^2 \bigg| {a_3} (\kappa {b_4}+{a_4} \omega )+{b_3} ({a_4} \kappa+{b_4} \omega )\bigg| ^2}{{p_3}+{p_4}}+\frac{({p_1}-{p_2})^2 \bigg| {b_2} {b_1} (2\omega )\bigg| ^2}{{p_1}+{p_2}}.
	\end{eqnarray}	
	
	$\bullet$ \textbf{Local quantum uncertainty}\\ 	
	
	Now, in order to obtain the explicit expression of local quantum uncertainty, we start by calculating the eigenvalues in Eq.(\ref{Symmetric matrix}). Indeed, they are given as: 
	\begin{eqnarray*}
		r_1&=&	\frac{4 \left| {\rho_{14}(t)}\right|  \left| {\rho_{23}(t)}\right| +({\rho_{22}(t)}-{\rho_{33}(t)}) ({\rho_{22}(t)}+{\rho_{33}(t)}+2 {\rho_{44}(t)}-1)}{\left(\sqrt{{p_1}}+\sqrt{{p_2}}\right) \left(\sqrt{{p_3}}+\sqrt{{p_4}}\right)}+\left(\sqrt{{p_1}}+\sqrt{{p_2}}\right) \left(\sqrt{{p_3}}+\sqrt{{p_4}}\right),\\
		r_2&=&	\frac{-4 \left| {\rho_{14}(t)}\right|  \left| {\rho_{23}(t)}\right| +({\rho_{22}(t)}-{\rho_{33}(t)}) ({\rho_{22}(t)}+{\rho_{33}(t)}+2 {\rho_{44}(t)}-1)}{\left(\sqrt{{p_1}}+\sqrt{{p_2}}\right) \left(\sqrt{{p_3}}+\sqrt{{p_4}}\right)}+\left(\sqrt{{p_1}}+\sqrt{{p_2}}\right) \left(\sqrt{{p_3}}+\sqrt{{p_4}}\right),\\
		r_3&=& \frac{1}{2} \left(\frac{({\rho_{22}(t)}+{\rho_{33}(t)}+2 {\rho_{44}(t)}-1)^2-4 |\rho_{14}(t)|^2}{\left(\sqrt{{p_1}}+\sqrt{{p_2}}\right)^2}+\frac{({\rho_{22}(t)}-{\rho_{33}(t)})^2-4 |\rho_{23}(t)|^2 }{\left(\sqrt{{p_3}}+\sqrt{{p_4}}\right)^2}\right),\\
		&+&\frac{1}{2} \left(\left(\sqrt{{p_1}}+\sqrt{{p_2}}\right)^2+\left(\sqrt{{p_3}}+\sqrt{{p_4}}\right)^2\right).
	\end{eqnarray*}
	Hence, the LQU is straightforwardly given as bellow: 
	\begin{equation}\label{LQU's formula}
		\mathcal{Q}(\rho_{Q}(t))=1-\max\{ r_1, r_2,r_3 \}. 
	\end{equation}
	
    \section*{Acknowledgment}
    During a visit to the Max Planck Institute for the Physics of Complex Systems, A.E.A. completed some of this work. He would like to thank the MPI-PKS for the financial support and friendly environment. The authors thank the reviewers for their insightful comments.

	\section*{Declaration of Interest}
	
	The authors declare that they have no conflict of interest.
	
	\section*{Data availability statement}
	
	No data statement is available.

\end{document}